\documentclass[%
 reprint,
superscriptaddress,
 amsmath,amssymb,
 aps,
prb,
]{revtex4-2}

\usepackage{graphicx}% Include figure files
\usepackage{dcolumn}% Align table columns on decimal point
\usepackage{bm}% bold math
\usepackage{xspace}
\usepackage{placeins}
\usepackage{xspace}
\usepackage{xcolor}

\renewcommand{\v}[1]{\ensuremath{\mathbf{#1}}}
\newcommand{\rtt}[1]{\textcolor{black}{#1}}
\begin{document}

\clearpage

\title{Critical role of magnetic moments on lattice dynamics in YBa$_2$Cu$_3$O$_6$}

\author{Jinliang Ning}
\affiliation{Department of Physics and Engineering Physics, Tulane University, New Orleans, Louisiana 70118, United States}

\author{Christopher Lane}
\affiliation{Theoretical Division and Center for Integrated Nanotechnologies, Los Alamos National Laboratory, Los Alamos, New Mexico 87545, USA}
\affiliation{Center for Integrated Nanotechnologies, Los Alamos National Laboratory, Los Alamos, New Mexico 87545, USA}

\author{Yubo Zhang}
\affiliation{MinJiang Collaborative Center for Theoretical Physics, College of Physics and Electronic Information Engineering, Minjiang University, Fuzhou 350108, China}

\author{Matthew Matzelle}
\affiliation{Department of Physics, Northeastern University, Boston, MA 02115, USA}

\author{Bahadur Singh}
\affiliation{Department of Condensed Matter Physics and Material Science, Tata Institute of Fundamental Research, Colaba, Mumbai 400005, India}

\author{Bernardo Barbiellini}
\affiliation{Department of Physics, School of Engineering Science, Lappeenranta University of Technology, FI-53851 Lappeenranta, Finland}
\affiliation{Department of Physics, Northeastern University, Boston, MA 02115, USA}

\author{Robert S. Markiewicz}
\affiliation{Department of Physics, Northeastern University, Boston, MA 02115, USA}

\author{Arun Bansil}
\affiliation{Department of Physics, Northeastern University, Boston, MA 02115, USA}

\author{Jianwei Sun}
\homepage{https://www.matcomp.org/}
\email{jsun@tulane.edu}
\affiliation{Department of Physics and Engineering Physics, Tulane University, New Orleans, Louisiana 70118, United States}

\date{\today}% It is always \today, today,

\begin{abstract}
The role of lattice dynamics in unconventional high-temperature superconductivity is still vigorously debated. Theoretical insights into this problem have long been prevented by the absence of an accurate first-principles description of the combined electronic, magnetic, and lattice degrees of freedom. Utilizing the recently constructed r$^2$SCAN density functional that stabilizes the antiferromagnetic (AFM) state of the pristine oxide YBa$_2$Cu$_3$O$_6$, we faithfully reproduce the experimental dispersion of key phonon modes. We further find significant magnetoelastic coupling in numerous high energy Cu-O bond stretching optical branches, where the AFM results improve over the soft non-magnetic phonon bands. 

%\begin{description}
%\end{description}
\end{abstract}

\maketitle

%\tableofcontents

\section{\label{sec:level1}Introduction}
Despite the discovery of unconventional high-temperature superconductivity in the cuprates a little over thirty years ago, there is still no consensus on the underlying microscopic mechanism\cite{1986HTc,RVBAnderson,pairRMP,arpsRMP,keimer2015,Sobota2021}. Early theoretical works \cite{bcsnotwork,giustino08nature,YBCO7LDA} suggested that conventional electron-phonon coupling does not play an important role in driving superconductivity in the cuprate family of materials. However, recent experiments find a more nuanced picture\cite{ELmixHTc,rezniknatureEPC,natureEPC,NCCOanomaly, LSCOanomaly,rezniknatureEPC,isotopeHTc,B2212EPC}. A strong anomaly in the Cu-O bond-stretching phonon beyond conventional theory is observed near optimal doping and is associated with charge inhomogeneity in the system\cite{rezniknatureEPC}. Optical spectroscopy reports find that antiferromagnetic spin fluctuations are the main mediators for the formation of Cooper pairs, but that the electron-phonon coupling gives a contribution to the bosonic glue of at least 10\%\cite{dal2012}. Moreover, recent ARPES observations suggest that the electronic interactions and the electron-phonon coupling reinforce each other in a positive-feedback loop, which in turn drives a stronger superconductivity\cite{B2212EPC}. 

One reason why the role of phonons was dismissed by the theoretical community is, in part, based on the failure of density functional theory (DFT) calculations to find any appreciable electron-phonon coupling at both the local density approximation (LDA) and generalized gradient approximation (GGA) levels\cite{giustino08nature}. This issue was further compounded by these density functional approximation's inability to stabilize the correct antiferromagnetic ground state in the parent phase, let alone its evolution with doping\cite{yubostripe,James_SCAN_Cuprates}. While corrections such as the addition of a Hubbard U \cite{HubbardU,linearU,LDAU_MnO,LSDAU_NiO} stabilize an antiferromagnetic (AFM) ground state \cite{sterling2021}, they can spoil the good agreement of PBE with experimental low-temperature equilibrium volumes\cite{jarlborg2014}. Therefore, it is evident that a {\it holistic} ab initio treatment is required to simultaneously satisfy both the lattice and magnetic degrees of freedom.

Recent advances in the construction of density functionals gives new hope in addressing the electronic structures of correlated materials at a first-principles level. Specifically, the strongly-constrained and appropriately-normed (SCAN) meta-GGA density functional \cite{SCAN,SCAN_NChem}, which satisfies 17 exact constraints, has demonstrated excellent performance across a diverse range of bonding environments. In particular, SCAN accurately predicts the correct half-filled AFM ground state and the transition from the insulating to the metallic state with doping observed in the cuprates \cite{yubostripe,James_SCAN_Cuprates}. Moreover, SCAN provides improved estimates of lattice constants, across correlated and transition metal compounds \cite{SCAN,SCAN_NChem, yubostripe, James_SCAN_Cuprates, yubo_MO, Yubo_TiO2, Peng_MnO2, Peng_MO, ning_MBT,SIO_Chris}. Thus, by accurately capturing the electronic and magnetic ground state SCAN has the potential to provide a good description of lattice dynamics and associated electron-phonon coupling. Unfortunately, SCAN suffers numerical problems that are exacerbated in phonon calculations, making reliably obtaining accurate phonon spectra from SCAN calculations a challenging task. Recently, we have shown that a revised version of SCAN, called r$^2$SCAN \cite{r2SCAN}, solves the numerical instability problem and delivers accurate, transferable, and reliable lattice dynamics for various systems with different bonding characteristics\cite{r2SCAN_phonon}.

In this article, we examine the role magnetoelastic effects play in explaining the experimental phonon dispersion of pristine YBa$_2$Cu$_3$O$_6$ by taking advantage of the numerically stable r$^2$SCAN functional. We find specific branches of the phonon band structure to be sensitive to the ground state magnetic order. Moreover, these phonons correspond to breathing modes within the CuO$_2$ plane, suggesting a sensitive dependence on magnetoelastic coupling, which may facilitate a positive-feedback loop between electronic, magnetic, and lattice degrees of freedom.

\section{\label{sec:level1}Methods}
{\it Ab initio} calculations were performed using the pseudopotential projector-augmented wave method\cite{PAW,PAW_vasp} implemented in the Vienna {\it ab initio} simulation package (VASP) \cite{VASP,VASP2} with an energy cutoff of 600 eV for the plane-wave basis set. Exchange-correlation effects were treated using the r$^2$SCAN \cite{r2SCAN, r2SCAN_phonon} meta-GGA scheme. The calculations are performed with a Gamma-centered mesh having a spacing threshold of 0.15 \AA$^{-1}$ for the $k$-space sampling. We used the experimental low-temperature P4/mmm crystal structure to initialize our computations\cite{ybco6_LC_expt}. All atomic sites in the unit cell along with the cell dimensions were relaxed using a conjugate gradient algorithm to minimize the energy with an atomic force tolerance of 0.001 eV/\AA~ and a total energy tolerance of $10^{-7}$ eV. The harmonic force constants were extracted from VASP using the finite displacement method (with displacement 0.015\AA) as implemented in the Phonopy code \cite{phonopy}. In some calculations, to give better agreement with the experimental volume an effective U was added to r$^2$SCAN\cite{r2SCAN}. 

\section{\label{sec:level1}Results}

\begin{figure*}[htb]
\centering
\includegraphics[width=\linewidth]{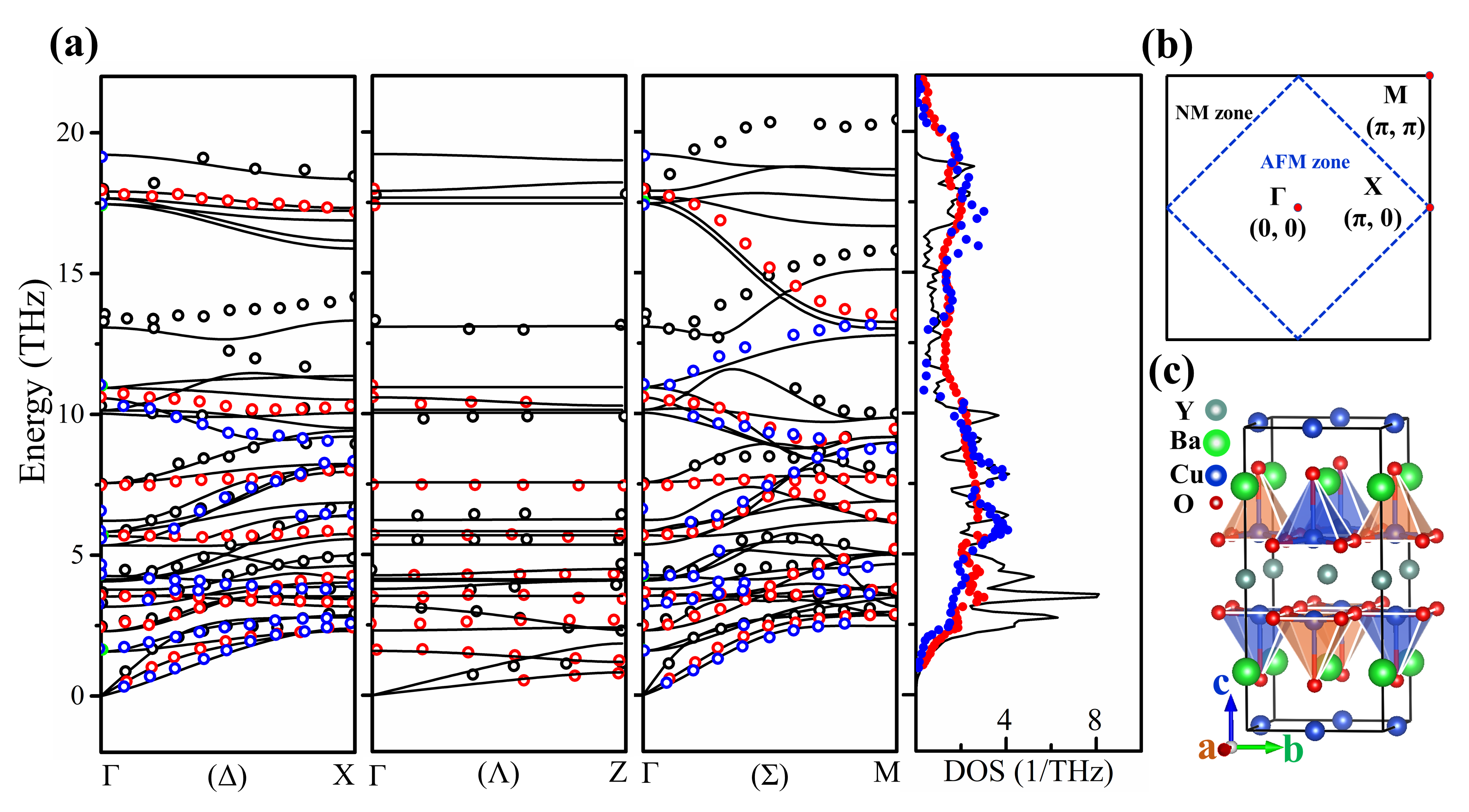}
\caption{(a) Theoretically predicted phonon band dispersions and density of states (DOS) of YBa$_2$Cu$_3$O$_6$ in the AFM phase (black lines) with the corresponding experimental data (circles for phonon data \cite{YBCO6phonon_expt}, and blue and red dots for DOS data measured at 78 K\cite{DOS1} and 6 K\cite{DOS2}, respectively). The irreducible representation of the various phonon modes, $\Delta_1$, $\Lambda_1$ and $\Sigma_1$ are denoted by black circles, $\Delta_2$ and $\Sigma_2$ by green circles, $\Delta_3$, $\Sigma_3$ and $\Lambda_5$ by red circles, and $\Delta_4$ and $\Sigma_4$ by blue circles, respectively. (b) A schematic of the NM (black dashed line) and AFM (blue dashed line) Brillouin zones; where the high-symmetry points used in (a) are marked. For simplicity, the high symmetry point Z ($0, 0, \pi$) is not shown. (c) $\sqrt{2}\times\sqrt{2}$ crystal structure of YBa$_2$Cu$_3$O$_6$ where the related AFM structure is highlighted by coloring the corner-sharing Cu-O pyramids blue(orange) for spin up(down)}\label{fig:phononbandsGAFMonly}
\end{figure*}

\begin{table*}[ht]
\caption{Calculated lattice constants, volume, Cu magnetic moment, Cu-O plane buckling angle $\angle$O-Cu-O, and Cu-O  $(d_{Cu-O})$ and Cu-O$_{ap}$ $(z^{(\prime)}_{Cu-O_{ap}})$ bond lengths for YBa$_2$Cu$_3$O$_6$ in non-magnetic and G-type AFM phases, along with the available experimental data. \rtt{Results from r$^2$SCAN+U (U = 1, 2, and 5 eV) and LDA+U (U = 5 and 8 eV) are included for comparison.}}
\label{tab:latticeU}
\begin{ruledtabular}
\begin{tabular}{llllllllll}
Methods & Phases &	$a$	(\AA)&	$c$ (\AA)	&	$V$	(\AA) &	$m (\mu_B)$	&	$d_{Cu-O}$ (\AA)	&	$\angle$O-Cu-O ($^\circ$)	&	$z_{Cu-O_{ap}}$ (\AA)	&	$z^{\prime}_{Cu-O_{ap}}$ (\AA)	\\
\cline{1-10}
	&	NM	&	3.8515	&	11.9890	&	177.85	&	0	&	1.934	&	169.25	&	1.804	&	2.576	\\
	&	AFM	&	3.8570	&	11.9417	&	177.65	&	0.45	&	1.937	&	169.28	&	1.805	&	2.554	\\
r$^2$SCAN	&	AFM(U=1)	&	3.8563	&	11.9281	&	177.38	&	0.50	&	1.937	&	168.89	&	1.803	&	2.543	\\
	&	AFM(U=2)	&	3.8561	&	11.9069	&	177.05	&	0.54	&	1.938	&	168.41	&	1.801	&	2.527	\\
	&	AFM(U=5)	&	3.8562	&	11.8321	&	175.95	&	0.66	&	1.941	&	167.00	&	1.795	&	2.472	\\
	\cline{1-10}
	&	NM	&	3.7930	&	11.6776	&	168.00	&	0	&	1.905	&	169.33	&	1.775	&	2.472	\\
LDA	&	AFM(U=5)	&	3.7839	&	11.5832	&	165.85	&	0.48	&	1.904	&	167.33	&	1.765	&	2.405	\\
	&	AFM(U=8)	&	3.7783	&	11.5188	&	164.44	&	0.59	&	1.903	&	166.17	&	1.755	&	2.363	\\
	\cline{1-10}
Expt.	&	AFM	&	3.8544$^a$	&	11.8175$^a$	&	175.57$^a$	&	0.55$^b$	&	1.940	&	166.78	&	1.786	&	2.471	\\

\end{tabular}
\end{ruledtabular}
\raggedright 
$^a$Powder neutron diffraction at temperature of 5 K\cite{ybco6_LC_expt}.\\
$^b$Single crystal neutron scattering\cite{casalta1994absence}.
\end{table*}

\begin{table*}[ht]
\caption{Calculated frequencies of dimpling, half-breathing, full-breathing, and buckling modes, in comparison with available experimental data. Calculated results include nonmangetic (NM) and AFM, as shown in Fig.~\ref{fig:phononbandsGAFMonly}, and AFM clculated by r$^2$SCAN+U (with U = 5 eV), as shown in Fig. 3.}
\label{tab:modes}
\begin{ruledtabular}
\begin{tabular}{lllllllllllll}
	&	\multicolumn{3}{c}{Dimpling (A)}			&			\multicolumn{3}{c}{Full-Breathing (G)}			&			\multicolumn{3}{c}{Half-Breathing (D)}			&			\multicolumn{3}{c}{Buckling (B)}		\\
	&	meV	&	cm$^{-1}$	&	THz	&	meV	&	cm$^{-1}$	&	THz	&	meV	&	cm$^{-1}$	&	THz	&	meV	&	cm$^{-1}$	&	THz	\\ \hline
NM	&	52.78	&	425.6	&	12.76	&	69.41	&	559.8	&	16.78	&	55.93	&	451.1	&	13.52	&	42.90	&	346.0	&	10.37	\\
AFM	&	54.14	&	436.6	&	13.09	&	76.35	&	615.8	&	18.46	&	71.67	&	578.0	&	17.33	&	41.48	&	334.6	&	10.03	\\
U5	&	57.04	&	460.0	&	13.79	&	81.97	&	661.1	&	19.82	&	75.03	&	605.1	&	18.14	&	42.35	&	341.6	&	10.24	\\
YBCO7 &	51.69	&	416.9	&	12.50	&	67.73	&	546.2	&	16.37	&	(X) 60.98	&	491.8	&	14.74	&	41.49	&	334.6	&	10.03	\\
      &		&		&		&		&		&		&	(Y) 55.53	&	447.8	&	13.42	&		&		&		\\
Expt.	&	56.08	&	452.3	&	13.56	&	84.50	&	681.5	&	20.43	&	76.31	&	615.4	&	18.45	&	42.60	&	343.6	&	10.30	\\
	&	(55.01	&	443.6	&	13.30)	&		&		&		&		&		&		&		&		&		\\

\end{tabular}
\end{ruledtabular}
\end{table*}

Table~\ref{tab:latticeU} compares various properties calculated with the non-magnetic (NM) and antiferromagnetic (AFM) states to available experimental values for YBa$_2$Cu$_3$O$_6$.  In the NM state, the lattice parameters and corresponding volume are the furthest away from the experimental values. Specifically, the in-plane lattice parameters slightly underestimate those from neutron scattering, whereas the predicted $c$-height is significantly larger, consistent with previous studies employing PBE and SCAN\cite{yubostripe,James_SCAN_Cuprates}. 

When the majority and minority spins are allowed to self-consistently relax, we stabilize the experimentally observed G-type AFM order across the planar copper atomic sites. The predicted value of the magnetic moment on copper sites is 0.45 $\mu_{B}$, which is in good accord with the corresponding experimental value of 0.55 $\mu_{B}$\cite{casalta1994absence}. Due to the localization of electrons on the copper sites the $ab$-plane expands with a concomitant shrinking of the c-axis, bringing the equilibrium crystal geometry in line with the experimental values. Since the phonon dispersion is determined by the inter-atomic forces, which depends sensitively on the ground state electronic structure and equilibrium atomic positions, the excellent performance of r$^2$SCAN in predicting the equilibrium ground state bodes well for an accurate prediction of the lattice dynamics.

\begin{figure*}[htb]
\centering
\includegraphics[width=\linewidth]{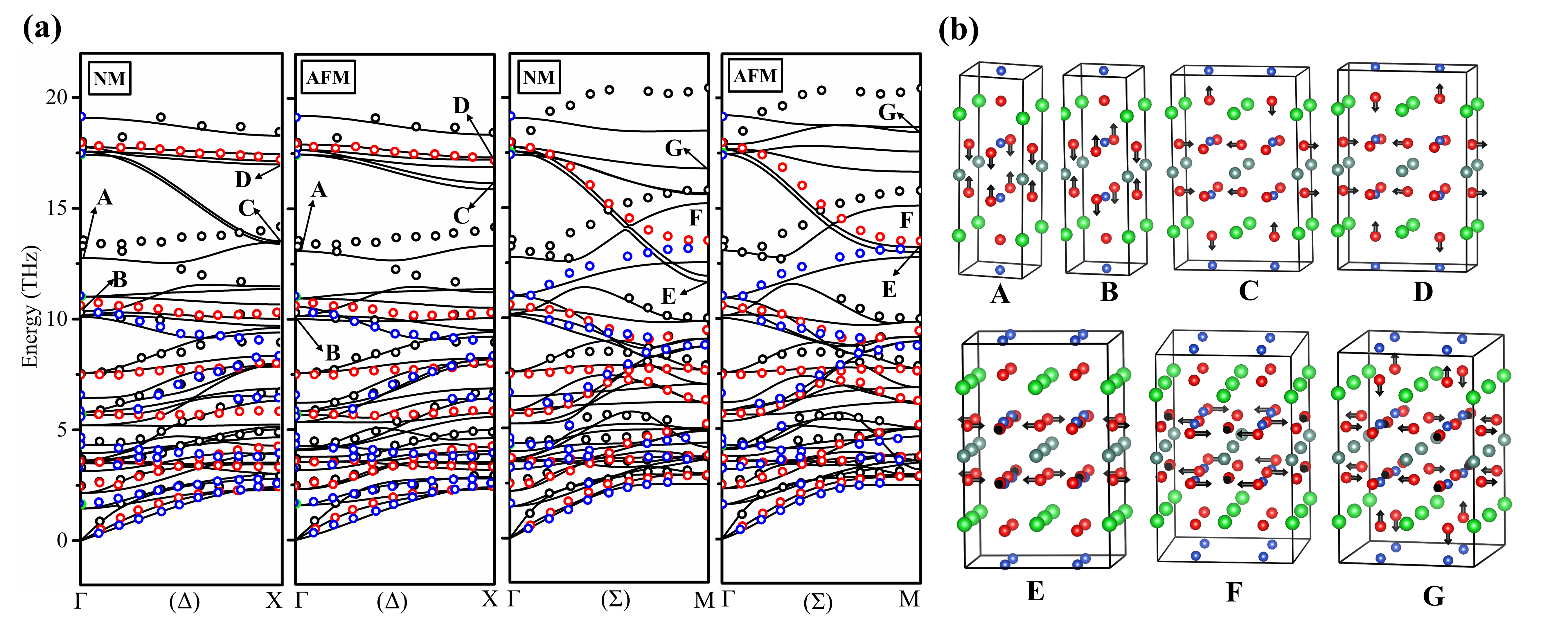}
\caption{(a) Comparison of phonon dispersions of YBa$_2$Cu$_3$O$_6$ in the nonmagnetic and and AFM phases (black lines)  with the corresponding experimental values\cite{YBCO6phonon_expt} (circles). The color scheme of the experimental bands is the same as in Fig.~\ref{fig:phononbandsGAFMonly}. (b) Schematic of the various vibrational modes most sensitive to the magnetic structure or/and experimentally interesting, as labeled in (a).}\label{fig:comparephononbands}
\end{figure*}

Figure~\ref{fig:phononbandsGAFMonly} (a) compares the theoretically predicted phonon dispersion of YBa$_2$Cu$_3$O$_6$ in the AFM phase with the experimental bands obtained by inelastic neutron scattering \cite{YBCO6phonon_expt, DOS1, DOS2}. For convenience we plotted the phonon spectra in the NM Brillouin zone of Fig.~\ref{fig:phononbandsGAFMonly} (b). Overall, r$^2$SCAN yields phonon frequencies and their dispersion along all three directions in momentum space in excellent accord with experiment.

To analyze the sensitivity of the phonon bands to magnetoelastic coupling we compare the NM and AFM phonon dispersion in Fig.~\ref{fig:comparephononbands} with the experimentally reported bands overlaid. By inspection, it is clear that a majority of the phonon branches are not significantly affected by the change in electronic environment [NM vs. AFM], which is consistent with the small difference in lattice constants given in Table~\ref{tab:latticeU}. However, for some branches, the AFM order results in a hardening of the affected phonon branches, i.e. a shift to higher frequency, which improves agreement with experiment. 

While overall agreement between experiment and the AFM phonon dispersions is quite good, for several branches there remains an underestimation of the hardness of the atomic bonds. Interestingly, many of these branches seem to be of relevance for the electronic properties of the cuprates. In Fig.~\ref{fig:comparephononbands}, we highlight seven branches, A-D along $\Gamma$-X, and E-G along $\Gamma$-M, for special discussion, illustrating the associated atomic displacements in Fig.~\ref{fig:comparephononbands}(b) and listing properties of the A, B, D, and G modes in Table~\ref{tab:modes}. Not only do these branches mostly show strong effects of magnetic order, but most are also of experimental interest, having strong doping dependence or changes at the superconducting transition. These branches are all related to the deformation (stretching or buckling) of Cu-O bonds. Branches A and B feature Cu-O bond out-of-plane buckling vibrations, while branch F features buckling in the CuO$_2$ plane. Branches C, D, E and G are all related to Cu-O bond stretching, whereas branches C, D and G also have a notable admixture of apical Cu-O bond stretching. This admixture has been noted in previous studies\cite{YBCO7_anomaly}. While branch D is usually called a half-breathing mode considering only the motion of the in-plane oxygens, there is a strong motion of the apical oxygen so we term this branch as an ac-plane full breathing mode. Branch G, normally called the full-breathing mode, is denoted as the 3D full-breathing mode. Mode E is normally called quadrupolar mode. Mode F is called the scissor mode in terms of the in-plane Cu-O bond buckling behavior. The Cu-O bond stretching modes are experimentally interesting due to their softening upon doping \cite{tachiki2003}, which relates to static charge density wave (CDW) or dynamic charge correlations. It is a distinct possibility that the dynamic charge correlations contribute to the superconducting pairing glue and/or are related to the formation of the pseudogap \cite{HBCO_DCC}. When YBa$_2$Cu$_3$O$_6$ is doped the half-breathing D branch evolves into the b-axis-polarized bond-stretching phonons of YBa$_2$Cu$_3$O$_7$, which gives a sharp local frequency minimum at wave vector \textbf{q}$\approx$0.27 upon cooling the sample from room temperature to T=10 K\cite{YBCO7_anomaly}. The B buckling mode also attracts interest because of its strong electron-phonon coupling and its possible connection to CDW formation\cite{CDW_buckling} and high-T$_c$ superconductivity \cite{YBCO6_EPC,EPCvsTc}. However, unlike the bond-stretching modes, this B$_1g$ bond-buckling branch does not change significantly between NM and AFM states. This could imply a different nature between the buckling and stretching modes in terms of their sensitivity to doping-induced changes in magnetism. 

For most of these bond-stretching modes, the strong magnetoelastic coupling effects associated with AFM order are sufficient to bring them into good agreement with experiment. Exceptions are the $ac$-plane and 3D full breathing branches (D and G), where the improvement from NM to AFM is not large enough. Note that $ac$-plane full-breathing branch D and the 3D full-breathing branch G correspond to the highest $\Delta1$ branch and the highest $\Sigma1$ branch, respectively, see Fig.~\ref{fig:phononbandsGAFMonly} (a). The underestimation of the D and G branches in frequency could result from the self-interaction error (SIE) in DFT that typically weakens bonding.

\section{\label{sec:level1}Discussion}
\begin{figure}[htb]
\centering
\includegraphics[width=\linewidth]{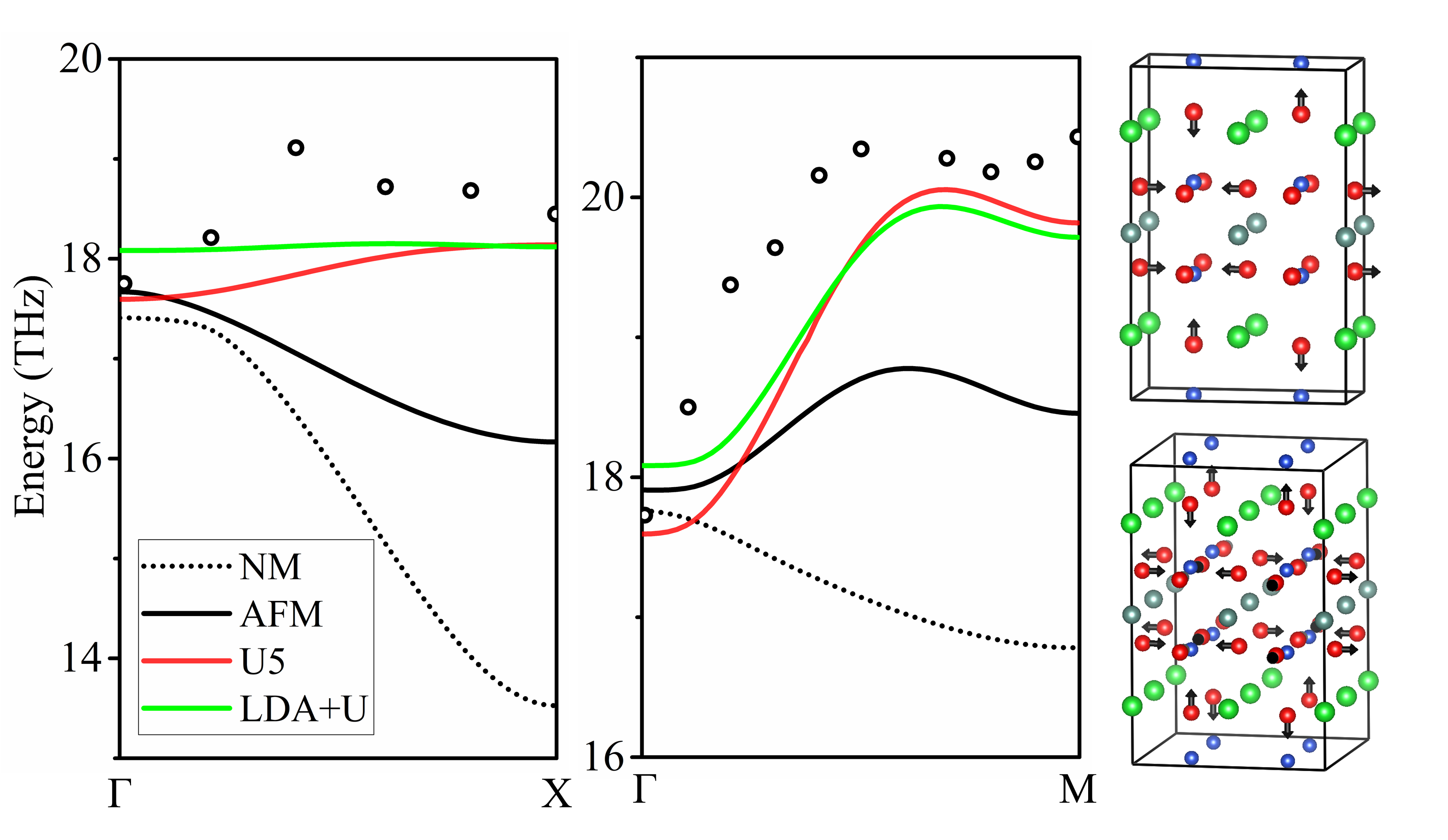}
\caption{Comparison of the calculated and experimental ac-plane full breathing branch (highest branch of $\Delta1$) and the 3D full breathing branch (highest branch of $\Sigma1$) for YBa$_2$Cu$_3$O$_6$. \rtt{Calculated results include the NM phase by r$^2$SCAN, and the G-type AFM phase by bare r$^2$SCAN, r$^2$SCAN+U with U = 5 eV (U5), and LDA+U with U = 8 eV.}}\label{fig:U}
\end{figure}
Due to the possible importance of the anomalous branches for superconductivity, we here briefly discuss possible explanations for the anomaly. While most branches in this energy range disperse downwards from $\Gamma$ towards X or M, branch A starts to disperse downward, but then all three branches disperse upwards. The plain r$^2$SCAN scheme without any U correction accurately captures the downward part of the dispersion, but then systematically underestimates the upward dispersion, with worst agreement for the highest energy branches D and G.  

In a practical DFT calculation, the exchange-correlation potential is always approximate, and it is commonplace that different realizations can have different performance on particular issues. Although SIE is reduced in SCAN-like functionals (r$^2$SCAN here), it still exists and may have significant impact in some situations, especially for strong-correlated systems like YBa$_2$Cu$_3$O$_6$. \rtt{In this work, we then test the effect of SIE in r$^2$SCAN on ac-plane and 3D full-breathing branches of YBa$_2$Cu$_3$O$_6$. For this purpose we employ the DFT+U scheme as implemented by Dudarev et al \cite{LSDAU_NiO}, which can effectively reduce SIE \cite{LDAU_MnO} of an underlying density functional. We caution the reader that there are a number of different treatments to reduce SIE (see Refs. \cite{LDAU_MnO,PZSIC,HSE,HSE06,Bernado_SIC} and references therein).} Table~\ref{tab:latticeU} shows that the best agreement with the experimental low-temperature equilibrium volume is actually obtained when a Hubbard U = 5 eV is added on the Copper sites in the AFM r$^2$SCAN  calculation. Figure~\ref{fig:U} shows that most of the discrepancies in {\it ac}-plane and 3D full-breathing branches can be cured by an additional magnetoelastic contribution explained by the same Hubbard U = 5 eV correction, although the transverse acoustic (lowest) phonon branches tend to become too soft, as shown in Fig.~\ref{fig:U5full}. This suggests that the phonon branches are sensitive to the anisotropy of the coulomb interaction. Note that this also increases the magnetic moment into better agreement with the experimental average. 

This finding is consistent with a recent LDA+U study on LaCu$_2$O$_4$ \cite{sterling2021}, where a large U correction (8 eV) is necessary for accurate calculation of the half- and full-breathing branches. Note our phonon results are self-consistently calculated from the fully relaxed geometry under the same r$^2$SCAN+U method, where the U correction improves the prediction on the lattice constants. \rtt{As shown in Table~\ref{tab:latticeU} and Figs. ~\ref{fig:U} and ~\ref{fig:U5full}, we added the geometry properties from LDA and LDA+U, and phonon dispersion results from LDA+U with U = 8 eV. The phonon dispersion from LDA+U is similar to that of r$^2$SCAN and r$^2$SCAN+U, but the lattice constants are worsened. LDA+U shows similar trend for lattice constants as r$^2$SCAN+U, namely, decreasing with increasing U. Since LDA already underestimates the lattice constants, adding U will worsen the lattice constant predictions for YBa$_2$Cu$_3$O$_6$. Therefore, LDA+U cannot give good predictions for geometry and phonon properties at the same time, while applying U on  r$^2$SCAN improves both. In some studies, people tend to use the experimental geometry for the LDA+U phonon dispersion calculations \cite{sterling2021}.} 

\rtt{In our recent work \cite{Kanun_NPJ}, we have discussed the crystal, electronic, and magnetic structures of La$_{2-x}$Sr$_x$CuO$_4$ for $x = 0.0$ and $x = 0.25$ employing 13 density functional approximations, representing the local, semi-local, and hybrid exchange-correlation approximations within the Perdew-Schmidt hierarchy. SCAN and r$^2$SCAN are found to perform well in capturing the key properties of La$_{2-x}$Sr$_x$CuO$_4$. In the future, a thorough comparative study will be conducted on lattice vibrational properties of typical cuprates, with special attention on performances of different density functional approximations and further improvements from corrections, such as Hubbard U and van der Waals interaction corrections.}
%LDA+U or GGA+U studies typically use experimental geometry for calculations of electronic structure and other properties, due to the concern that the semi-empirical U correction often degrades predictions on crystal structures. 

%The compatibility between U correction and the bare r$^2$SCAN functional might provide a self-consistent treatment of materials properties with SIC.

\section{\label{sec:level1}Conclusions}
In summary, we find that r$^2$SCAN's improved ability to treat electronic, magnetic, and structural correlations on equal footing carries over into the realm of lattice dynamics. Specifically, we achieve good agreement between experiment and theory for the phonon spectra of an insulating cuprate, YBa$_2$Cu$_3$O$_6$. Furthermore, by comparing the calculated phonon dispersions of the nonmagnetic and the AFM phases of YBa$_2$Cu$_3$O$_6$, we are able to show that strong magnetoelastic effects are crucial in reproducing the experimental results for the Cu-O bond stretching modes.

A notable residual disagreement in the {\it ac}-plane and 3D full breathing branches may be significant, as the modes involved strongly coupling to electrons.  By applying the r$^2$SCAN+U method, we achieved further improvements in both crystal geometry and these challenging phonon branches. This success of the r$^2$SCAN+U method provides a holistic description, where charge, magnetism, and lattice dynamics are treated on the same footing. Recent research efforts have renewed interest in the role of electron-phonon coupling in the mechanism of high-temperature superconductivity in cuprates \cite{Ahmadova2020}. Therefore, the fact that DFT is now capable of describing the electronic structure and lattice dynamics accurately at a fundamental level, paves the way for further investigation including phonon anomalies, charge inhomogeneity, cavity-phonon-magnon quasiparticle interactions \cite{cavity-phonon-magnon}, and phase competition, which in turn will provide unique insight into the high temperature superconducting materials. \rtt{It will be interesting to see if r$^2$SCAN or r$^2$SCAN+U modifies the YBa$_2$Cu$_3$O$_7$ electron-phonon coupling \cite{bcsnotwork,giustino08nature,YBCO7LDA} enough to enhance role of phonons in high-temperature superconductivity in cuprates.}

\begin{acknowledgments}
J.N. and J.S. acknowledge the support of the U.S. Office of Naval Research (ONR) Grant No. N00014-22-1-2673. Computational work done at Tulane University was supported by the Cypress Computational Cluster at Tulane and the National Energy Research Scientific Computing Center. The work at Northeastern University was supported by the US Department of Energy (DOE), Office of Science, Basic Energy Sciences Grant No. DE-SC0022216 (modeling complex magnetic states in materials). The work at Los Alamos National Laboratory was supported by the U.S. DOE NNSA under Contract No. 89233218CNA000001 and by the Center for Integrated Nanotechnologies, a DOE BES user facility, in partnership with the LANL Institutional Computing Program for computational resources. Additional support was provided by DOE Office of Basic Energy Sciences Program E3B5. The work at TIFR Mumbai was supported by the Department of Atomic Energy of the Government of India under Project No. 12-R\&D-TFR-5.10-0100. B.B. was supported by the Ministry of Education and Culture (Finland).
\end{acknowledgments}

%\clearpage
\onecolumngrid
\appendix*
\section{Additional data sets and figures}

The figure in this Appendix expands on the data presented in the main text. Figure 4 includes the full set of data from \rtt{r$^2$SCAN+U and LDA+U}, in comparison with the bare r$^2$SCAN results and experimental data.

\begin{figure}[htb]
\centering
\includegraphics[width=\linewidth]{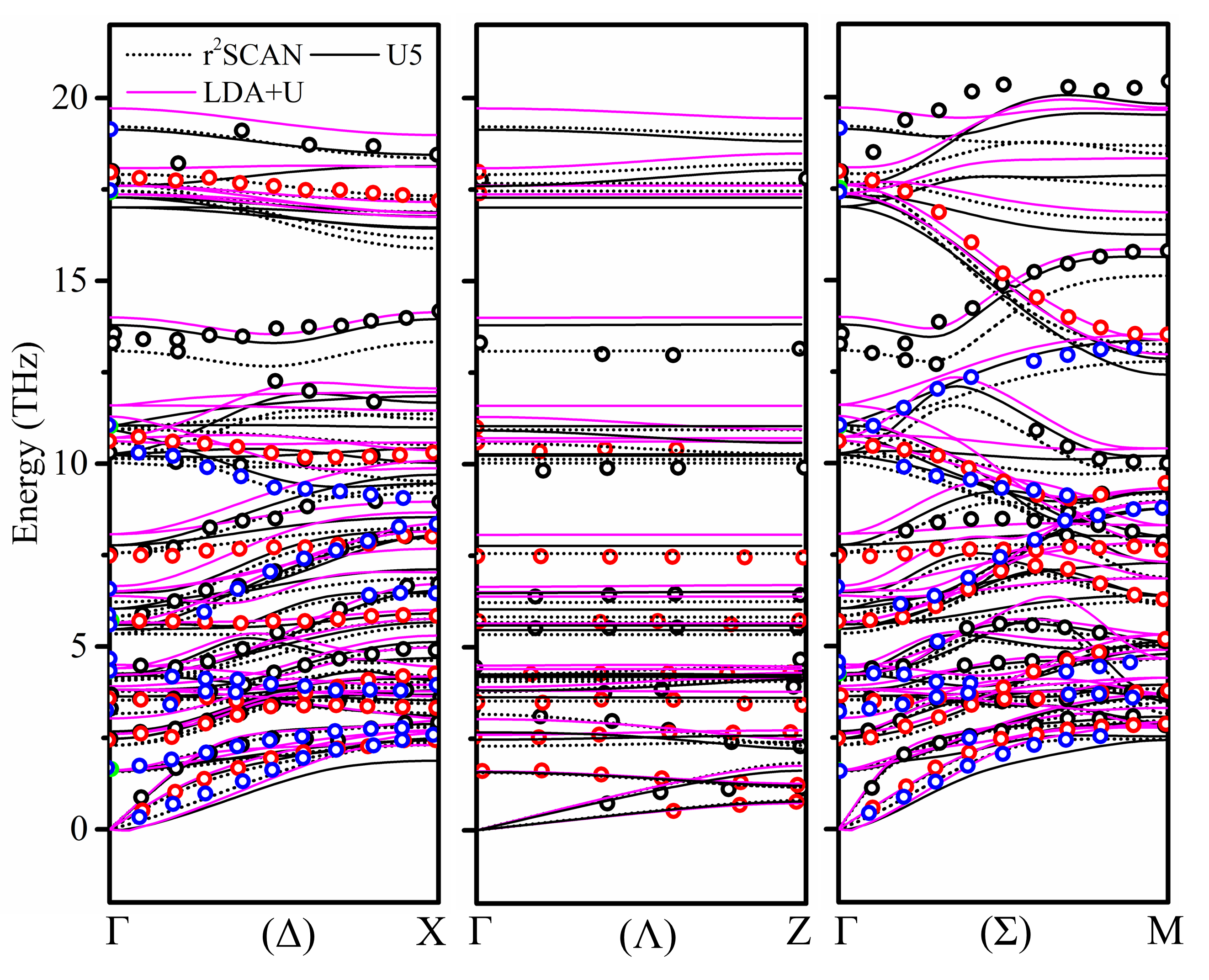}
\caption{Comparison of the calculated and experimental phonon dispersion for YBa$_2$Cu$_3$O$_6$. \rtt{Calculated results are for the G-type AFM phase by bare r$^2$SCAN, r$^2$SCAN+U with U = 5 eV (U5), and LDA+U with U = 8 eV.}}\label{fig:U5full}
\end{figure}

\twocolumngrid
\bibliography{ref}% Produces the bibliography via BibTeX.

\end{document}